\documentclass[%
 aip,
 amsmath,amssymb,
 reprint,%
nofootinbib,
]{revtex4-1}

\usepackage{graphicx}% Include figure files
\usepackage{dcolumn}% Align table columns on decimal point
\usepackage{bm}% bold math
\usepackage[utf8]{inputenc}
\usepackage[T1]{fontenc}
\usepackage{mathptmx}
\usepackage{etoolbox}
\usepackage{graphicx}% Include figure files
\usepackage{dcolumn}% Align table columns on decimal point
\usepackage{bm}% bold math
\usepackage{hyperref}% add hypertext capabilities
\usepackage{physics}
\usepackage{siunitx}
\usepackage[dvipsnames]{xcolor}

\usepackage{soul}

\makeatletter
\def\@email#1#2{%
 \endgroup
 \patchcmd{\titleblock@produce}
  {\frontmatter@RRAPformat}
  {\frontmatter@RRAPformat{\produce@RRAP{*#1\href{mailto:#2}{#2}}}\frontmatter@RRAPformat}
  {}{}
}%
\makeatother

\begin{document}

\preprint{AIP/123-QED}

\title{A perspective on \textit{ab initio} modeling of polaritonic chemistry: %and t
The role of non-equilibrium effects and quantum collectivity %for groundstate polaritonic reactions }%phenomena}
}
% Force line breaks with \\
\author{Dominik Sidler}
  \email{dsidler@mpsd.mpg.de}
 \affiliation{Max Planck Institute for the Structure and Dynamics of Matter and Center for Free-Electron Laser Science, Luruper Chaussee 149, 22761 Hamburg, Germany}
\affiliation{The Hamburg Center for Ultrafast Imaging, Luruper Chaussee 149, 22761 Hamburg, Germany}%Lines break automatically or can be forced with \\

\author{Michael Ruggenthaler}
  \email{michael.ruggenthaler@mpsd.mpg.de}
  \affiliation{Max Planck Institute for the Structure and Dynamics of Matter and Center for Free-Electron Laser Science, Luruper Chaussee 149, 22761 Hamburg, Germany}
  \affiliation{The Hamburg Center for Ultrafast Imaging, Luruper Chaussee 149, 22761 Hamburg, Germany}
  \author{Christian Sch\"afer}%
\email{christian.schaefer.physics@gmail.com}
\affiliation{Max Planck Institute for the Structure and Dynamics of Matter and Center for Free-Electron Laser Science, Luruper Chaussee 149, 22761 Hamburg, Germany}
\affiliation{The Hamburg Center for Ultrafast Imaging, Luruper Chaussee 149, 22761 Hamburg, Germany}
\affiliation{Department of Physics, Chalmers University of Technology, 412 96 G\"oteborg, Sweden}
\affiliation{Department of Microtechnology and Nanoscience, MC2, Chalmers University of Technology, 412 96 G\"oteborg, Sweden}
\author{Enrico Ronca}
  \email{enrico.ronca@pi.ipcf.cnr.it}
  \affiliation{Istituto per i Processi Chimico Fisici del CNR (IPCF-CNR), Via G. Moruzzi, 1, 56124, Pisa, Italy}
\author{Angel Rubio}
  \email{angel.rubio@mpsd.mpg.de}
  \affiliation{Max Planck Institute for the Structure and Dynamics of Matter and Center for Free-Electron Laser Science, Luruper Chaussee 149, 22761 Hamburg, Germany}
    \affiliation{The Hamburg Center for Ultrafast Imaging, Luruper Chaussee 149, 22761 Hamburg, Germany}
  \affiliation{Center for Computational Quantum Physics, Flatiron Institute, 162 5th Avenue, New York, NY 10010, USA}
  \affiliation{Nano-Bio Spectroscopy Group, University of the Basque Country (UPV/EHU), 20018 San Sebasti\'an, Spain}

\date{\today}% It is always \today, today,

\begin{abstract}
This perspective provides a brief introduction into the theoretical complexity of polaritonic chemistry, which emerges from the hybrid nature of strongly coupled light-matter states. To tackle this complexity, the importance of \textit{ab initio} methods is highlighted.
Based on those, novel ideas and research avenues are developed with respect to quantum collectivity, as well as for resonance phenomena immanent in reaction rates under vibrational strong coupling. 
Indeed, fundamental theoretical questions arise about the mesoscopic scale of quantum-collectively coupled molecules, when considering the depolarization shift in the interpretation of experimental data.
Furthermore, to rationalise recent findings based on quantum electrodynamical density-functional theory (QEDFT), a simple, but computationally efficient, Langevin framework is proposed, based on well-established methods from molecular dynamics. It suggests the emergence of cavity induced non-equilibrium nuclear dynamics, where thermal (stochastic) resonance phenomena could emerge in the absence of external periodic driving. Overall, we believe the latest \textit{ab initio}  results indeed suggest a paradigmatic shift for  ground-state chemical reactions under vibrational strong coupling, from the  collective quantum interpretation towards a more local, (semi)-classically and non-equilibrium dominated perspective. 
Finally, various extensions towards a refined description of cavity-modified chemistry are introduced in the context of QEDFT and future directions of the field are sketched.

\end{abstract}

\maketitle

\section{Introduction}

Polaritonic chemistry has become a rapidly developing field over the last years, driven by numerous experimental breakthroughs, which nurture the hope for unprecedented (quantum) control in chemistry. For example, experimental realizations confirmed that vibrational strong coupling can inhibit,\cite{thomas2016ground} steer,\cite{thomas2019tilting} and even catalyze\cite{hiura2018cavity} a chemical process. Moreover, seminal measurements were published on the control of photo-chemical reactions,\cite{munkhbat2018suppression} energy transfer,\cite{coles2014polariton} the realization of single molecular strong coupling\cite{wang2017coherent} and even evidence for the increase of the critical temperature in superconductors was reported.\cite{thomas2019exploring}

In parallel with these outstanding experimental efforts, the development of theoretical methods flourished, aiming for the detailed understanding of the underlying driving mechanism of polaritonic chemistry. However, the emergence of hybrid light-matter states poses a notoriously hard problem to capture theoretically.\cite{ruggenthaler2018quantum} Aside from the generally well known complexity of the electron-nuclei dynamics under variable chemical conditions, %(e.g. solvent effects or different equilibrium ensembles)
 strong coupling to the electromagnetic field introduces fundamentally new (quantum) states, i.e. polaritons, which give rise to a dramatic increase in chemical and computational complexity, due to the large dimensionality of the combined light-matter degrees of freedom.
For example, the emergence of collective coupling effects can transfer energy over distances $\geq 100$ nm.\cite{zhong2017energy} At the same time, collective coupling is also believed to introduce quantum coherence on a mesoscopic scale at ambient conditions,\cite{guebrou2012coherent} which mitigates the locality assumption prevalent in chemistry.  Moreover, strong light matter interaction leads to the formation of correlated  dark states,\cite{gonzalez2016uncoupled} i.e. excitations that cannot be populated by the absorption of light, which  boost the chemical complexity even further. %Additionally, the strong coupling to multiple photon modes may alter emission and dissipative properties of the polaritonic system,[?] which eventually modifies lifetimes or reaction rates  governing the chemical system under investigation.
Overall, this astonishing diversity of polaritonic chemistry opens a plethora of novel perspectives on tailoring chemistry\cite{ebbesen2016hybrid} or designing novel materials,\cite{hubener2021engineering} and it even leaves room for fundamental new discoveries such as novel phases.\cite{thomas2021large,latini2021phonoritons,latini2021ferroelectric} To account for this vast complexity theoretically, computational methods have been developed over the past years, which range from phenomenologically driven approaches in quantum optics \cite{jaynes1963comparison,herrera2016cavity,ribeiro2018polariton} over semi-classical descriptions \cite{li2020cavity} and properly designed orbital theories\cite{RisoNatureCommun2022} up to the full \textit{ab initio} setting in non-relativistic quantum electrodynamics (QED).\cite{ruggenthaler2014quantum,flick2018ab,haugland2020coupled} However, despite this broad range of theoretical methods, there is no consensus in the field about the necessary and sufficient conditions to apply the different methodologies and fundamental theoretical questions remain open.
The goal of the following perspective  is to illustrate these fundamental problems and opportunities from an \textit{ab initio} perspective. For this purpose a brief introduction to the theoretical foundations of \textit{ab initio} QED with a focus on quantum electrodynamical density-functional theory (QEDFT) is given and the unique benefits are illustrated for realistic polaritonic settings. Application-wise, we focus mainly on cavity-assisted reaction dynamics to scrutinize common theoretical assumptions under vibrational strong coupling, such as the emergence of mesoscopic, collective quantum states. Our considerations suggest a more localized, semi-classical perspective of polaritonic chemistry, where the theoretically elusive resonance condition emerges due to cavity-induced non-equilibrium effects. Finally, extensions to a more complete description of cavity-modified chemistry beyond the current state-of-the-art are discussed and a road-map of future developments in \textit{ab initio} QED and polaritonic chemistry is sketched.

\section{State-of-the-art theoretical description of polaritonic chemistry} 
\subsection{\textit{Ab initio} theory and its relation to phenomenological models}

A priori the strong hybridization of light and matter requires a non-perturbative, self-consistent treatment of light and matter at relativistic scales by means of quantum electrodynamics (QED). However, so far this most accurate theoretical description available is only applicable perturbatively to scattering processes. This limits its feasibility for the highly non-perturbative processes involved in chemical reactions, e.g., when the structure of a molecule is considerably changed. However, when going to the non-relativistic limit, the fundamental drawback of QED is lifted and it provides access to a self-consistent description of polaritonic processes by solving the Schrödinger equation for the Pauli-Fierz Hamiltonian $\hat{H}$.\cite{spohn2004dynamics,craig1998molecular,schaefer2020relevance} Here, the Pauli-Fierz Hamiltonian is introduced in the long-wavelength limit in the length gauge  for interacting matter strongly coupled to $M$ cavity modes $\alpha$, which is the fundamental ingredient of recent state-of-the-art \textit{ab initio} methods in polaritonic chemistry. 
\begin{eqnarray}
\hat{H} &=&\sum_i^n \frac{\hat{\bold{p}}_i^2}{2m}+\sum_i^N \frac{\hat{\bold{P}}_i^2}{2M_i}+\sum_{i<j}^n\frac{e^2}{|\hat{\bold{r}}_i-\hat{\bold{r}}_j|}+\sum_{i<j}^N\frac{e^2 Z_i Z_j}{|\hat{\bold{R}}_i-\hat{\bold{R}}_j|}\nonumber\\
&&-\sum_{i,j}^{n,N}\frac{e^2 Z_j}{|\hat{\bold{r}}_i-\hat{\bold{R}}_j|}
 +\sum_\alpha^M\frac{1}{2}\bigg[\hat{p}_\alpha^2+\omega_\alpha^2\Big(\hat{q}_\alpha-\frac{\boldsymbol{\lambda}_\alpha}{\omega_\alpha}\cdot \hat{\bold{X}}\Big)^2\bigg]
%+ \bold{E}_{\mathrm{ext}}\cdot \hat{\bold{X}}+\sum_\alpha^M \frac{j_{\mathrm{ext}}^\alpha}{\omega_\alpha}\hat{q}_\alpha\
% &=&\sum_i^n \frac{\hat{\bold{p}}_i^2}{2m}+\sum_i^N \frac{\hat{\bold{P}}_i^2}{2m_i}+\sum_{i<j}^n\frac{e^2}{|\hat{\bold{r}}_i-\hat{\bold{r}}_j|}+\sum_{i<j}^N\frac{e^2 Z_i Z_j}{|\hat{\bold{R}}_i-\hat{\bold{R}}_j|}\nonumber\\
%&&+\sum_{i,j}^{n,N}\frac{e^2 Z_j}{|\hat{\bold{r}}_i-\hat{\bold{R}}_j|}
% +\sum_\alpha^M\frac{1}{2}\bigg[\hat{p}_\alpha^2+\omega_\alpha\Big(\hat{q}_\alpha-\frac{\boldsymbol{\lambda}_\alpha}{\omega_\alpha}\cdot \hat{\bold{X}}\Big)^2\bigg]
%+ \bold{E}_{\mathrm{ext}}\cdot \hat{\bold{X}}+\sum_\alpha^M \frac{j_{\mathrm{ext}}^\alpha}{\omega_\alpha}\hat{q}_\alpha\
\label{eq:pf_dip_h}
\end{eqnarray}

All $n$ electrons and $N$ nuclei interact via the Coulomb interaction assuming atomic units. The unit particle mass is indicated by $m=1$ to distinguish from the nuclear mass $M_i$ and the unit charge is given by $e=1$ with nuclear charge number $Z_i$. The canonical position $\hat{\textbf{r}}_i$ and momentum $\hat{\textbf{p}}_i$ operators are defined, where small letters indicate electrons and capital letters the nuclei. The canonical displacement field operators of the photon field are given by $\hat{q}_\alpha$, $\hat{p}_\alpha$. The mode frequency is labeled by $\omega_\alpha$ and the light-matter coupling by $\boldsymbol\lambda_\alpha$, which inversely depends on the mode volume. The total dipole operator of electrons and nuclei is defined by $\hat{\bold{X}}$.

Notice that the difficulty of the eigenvalue problem imposed by Eq. (\ref{eq:pf_dip_h}) is beyond quantum mechanics, even in the long-wavelength limit. Hence, the exact solution of the Pauli-Fierz Hamiltonian is completely intractable except for low-dimensional model systems or as recently shown  for 3-body quantum systems (He, HD+ or H2+) coupled to a single photon mode.\cite{sidler2020chemistry}  Due to this complexity, it has become a common standard in the field of polaritonic chemistry to circumvent this issue by combining simple phenomenological model Hamiltonians from quantum optics (e.g. Dicke\cite{dicke1954coherence} or 
Jaynes-Cummings\cite{jaynes1963comparison} model), with various standard methods from computational chemistry (e.g. density functional theory (DFT),\cite{galego2015cavity} molecular dynamics (MD),\cite{luk2017multiscale} surface hopping\cite{fregoni2018manipulating}). Often these phenomenological approaches provide a very intuitive insight into polaritonic processes with relatively little additional computational costs. Consequently, they gathered large popularity and are widespread among the scientific community. For example, parametrized phenomenological models have been proven successful in reproducing spectral observables.\cite{f2018theory} In addition, they provide a powerful approach to include dissipative processes by means of Lindblad terms or to extrapolate to large system sizes, far beyond any explicit computational description, in the dilute gas limit.\cite{reitz2019langevin} However, despite this impressive success, cases have been reported, where model predictions contradict experimental observations,\cite{thomas2016ground,thomas2020comment} which has triggered controversial discussions between theoreticians and experimentalists.\cite{climent2020sn,thomas2020comment,climent2021reply} Overall, there is a general consensus among the community that we still lack a detailed theoretical understanding of the relevant processes involved in polaritonic chemistry and considerable research effort is needed to unravel them. Overcoming this theoretical shortcoming is of eminent importance for the maturity of the entire field. Eventually, one desires a level of understanding that can boost the development of future industrial applications not only experimentally, but also theoretically. 
From the authors' perspective, the route towards a general consensus between experiment and theory can be separated into two distinguishable theoretical branches: 
\begin{enumerate}
    \item The continuous refinement of existing phenomenological model based approaches remains of great importance. If applied correctly, a suitable model ideally allows the direct study of the underlying driving mechanisms, which is crucial to get a physical intuition of polaritonic chemistry. 
    \item The rigorous theoretical description based on the full non-relativistic Pauli-Fierz Hamiltonian is required, where all additionally involved approximations and assumptions are well defined and can be relaxed if necessary. Such a rigorous \textit{ab initio} method is vital to benchmark aforementioned models and it is the only way to generate unbiased and reliable theoretical insight beyond the predetermined intuition of a model Hamiltonian. 
\end{enumerate}

 Over the last years, considerable research effort has been invested into the development of the \textit{ab initio} research branch. This has culminated in the introduction of QEDFT\cite{ruggenthaler2014quantum,flick2018ab} and even polaritonic coupled cluster\cite{haugland2020coupled} methods that are applicable to realistic chemical setups.  While the versatile QEDFT (see also discussion in Sec.~\ref{sec:FutureQEDFT}) provides an optimal balance between accuracy and computational efficiency, coupled cluster methods give even access to the accurate study of polaritonic quantum correlations.

\begin{figure*}
\includegraphics[width=\textwidth]{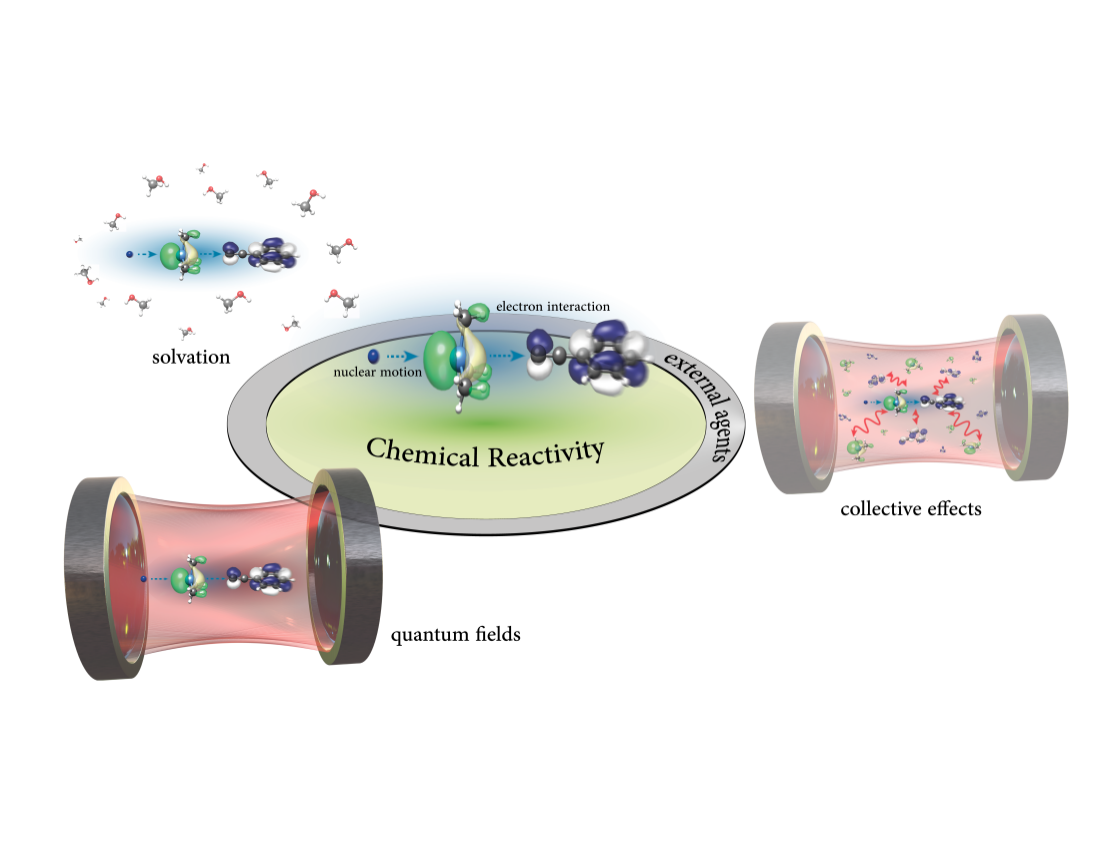}% Here is how to import EPS art
\caption{\label{fig:wide} Illustration of the main sources of complexity for polaritonic chemistry. Besides the well-known complexity of the chemical systems themselves (indicated by \textit{chemical reactivity}), and the influence of \textit{solvation}, we also have an increase of complexity due to the \textit{quantum fields} and the the \textit{collective effects} that can arise when many molecules undergo a reaction inside an optical cavity.}
\end{figure*}

\subsection{Towards unravelling the mystery of cavity-mediated reaction rates under vibrational strong coupling\label{sec:list}}

After the brief introduction to the  methodical aspects of polaritonic chemistry, we will focus more on the physical origin of the observed experimental results. Numerous potentially relevant effects arise, due to the vast complexity of polaritonic systems.  The most prominent sources of complexity are illustrated in Fig.~1 and include the complexity of the chemical systems itself, the complexity due to  solvation, the complexity due to potential quantum effects of the radiation field and the complexity due to a large ensemble of molecules. Several fundamental theoretical questions emerge, such as:
%Despite the impressive experimental advances to control chemical reaction rates,\cite{?} we lack a detailed understanding of the complex interplay of multiple relevant mechanisms and aspects, which for example involve:

\begin{enumerate}

    \item Are local or non-local effects dominant (e.g. charge vs. energy transfer)?
    \item Can the effect be captured classically?
    \item Arises the effect due to collective or single-molecule strong coupling?
    \item Is it an equilibrium or a non-equilibrium effect?
    \item Is only one mode of the cavity relevant or are its multi-mode character and losses important?
    \item Are classical correlations or quantum entanglement important in collective coupling?
    %\item The impact of lossy cavities.
    \item To what extend are spatial variations of the cavity-modes important (beyond dipole coupling or spin interactions)?
    \item Can a theoretical method that describes one specific constituent (electronic, nuclear and photonic degrees of freedom) or observable (e.g. Rabi splitting) be applied to  different constituents or different observables (e.g. chemical reaction)?

    %\item The connection of different observables, e.g. what can be inferred about cavity mediated reaction rates by meeting the strong coupling conditions determined from absorption measurements (Rabi-splitting)?
    %\item The origin and role of different resonance phenomena, e.g. externally driven absorption measurements (Rabi-splitting)  vs. internally driven (?) resonance properties of chemical reaction rates.
\end{enumerate}

Certainly, determining the decisive mechanisms for cavity mediated reactions among these aspects will strongly depend on the chemical system under study and the chosen cavity setup, i.e.:
\begin{enumerate}
    \item Do the cavity photons couple strongly to electrons or nuclei (electronic vs. vibrational coupling)?
    \item Is it a ground-state or excited-state reaction (e.g. electron transfer vs. photo-chemical reactions)?
    \item Does the cavity couple strongly to the solute or the solvent molecules or both? What is the impact of the state of matter under study? For example, does the reaction occur in the gaseous or in the liquid phase, or are there even solids involved as  catalysts. 
    \item What experimental cavity realization is chosen, which determines the collective and local light-matter coupling?
    %\item The connection of different observables, e.g. what can be inferred about cavity mediated reaction rates by meeting the strong coupling conditions determined from absorption measurements (Rabi-splitting)?
    %\item The origin and role of different resonance phenomena, e.g. externally driven absorption measurements (Rabi-splitting)  vs. internally driven (?) resonance properties of chemical reaction rates.
\end{enumerate}
From these lists it becomes immediately clear that a detailed understanding and theoretical description of cavity-mediated processes is a highly non-trivial problem. 
Categorizing and disentangling these effects to reach an intuitive understanding of polaritonic chemistry can probably be considered as the major goal of the entire polaritonic community.
To our opinion, recently developed \textit{ab initio} methods (e.g. QEDFT) provide a mostly unbiased approach to tackle this enormous complexity with as little preliminary assumptions and restrictions as possible. These insights  combined with experimental results can be used to advance our understanding of photon-modified chemical reactions. In the following we analyze a prototypical experiment, highlight possible inconsistencies that arise when applying common models to the problem and propose, based on the simplest practical model of chemical reactions, a local and mostly classical perspective that can serve as a computationally feasible starting point for future investigations. %In the following, we  aim to combine recently obtained  \textit{ab-initio} and experimental results with a logical argument to develop a concise hypothesis about the main mechanisms in cavity-mediated reaction dynamics under vibrational strong coupling.

\subsubsection{Resonance phenomena in cavity-mediated reaction rates under vibrational strong coupling}

The seminal experimental results of Ebbesen's group about the inhibition of the deprotection reaction of 1-phenyl-2-trimethylsilylacetylene (PTA) under vibrational strong coupling\cite{thomas2016ground} is the starting point of our subsequent theoretical arguments. Experimental evidence reveals an intriguing feature of cavity-mediated reaction rates. It shows that  tuning the cavity in resonance with a specific vibration is a crucial ingredient to lower the reaction rates.\cite{schafer2021shining} This prototypical result has triggered a controversial discussion between theoreticians\cite{climent2020sn,climent2021reply} and experimentalists \cite{thomas2016ground,thomas2020comment}  on the interpretation of the experimental results.  The existence of such a subtle resonance condition could not be predicted from equilibrium transition-state theory\cite{galego2019cavity,campos2020polaritonic,li2020origin} and a frequency dependency could only be predicted by dynamical solvent caging effects,\cite{li2021cavity} when tuning the cavity resonant with respect to the curvature of the potential energy surface at the transition state. \cite{li2021cavity,schafer2021shining} Recent \textit{ab initio} simulations could indeed confirm the existence of a dynamical caging effect at $\omega_c\approx 86$cm$^{-1}$ for the dressed PTA reaction.\cite{schafer2021shining} However, the computed resonance frequency is far below the experimentally observed resonance at  $\omega_c= 860$cm$^{-1}$,\cite{thomas2016ground} and remains inaccessible with today's experimental setups.
This suggests that the experimentally observed resonance phenomena relies on a different physical mechanism, which indicates that present phenomenological models cannot capture all relevant aspects in a polaritonic setting.\cite{simpkins2021mode}  
Indeed, very recently, \textit{ab initio} simulations based on Ehrenfest dynamics revealed novel aspects in this polaritonic reaction rate mystery. They uncovered that the presence of a cavity correlates different vibrational degrees of freedom in the investigated PTA complex, which effectively redistributes kinetic energy from a specific bond to other degrees of freedom, eventually causing the suppression of the bond breaking.\cite{schafer2021shining} This delicate dynamic redistribution proves to be sensitive with respect to the chosen resonance frequency,\cite{schafer2021shining} which is in qualitative agreement with experimental evidence.\cite{thomas2016ground,simpkins2021mode}
However, comparison between theory and experiment is further complicated by the yet unclear interplay between collective and local light-matter interaction, which can be expected to affect the sensitivity of the resonance condition. For example, the fundamental condition to reach strong coupling demands a sufficient (collective) oscillator strength to overcome the decoherence that will overshadow any hybridization between light and matter. Naturally, the oscillator strength is sensitive to the resonant condition, i.e., only if matter and photonic excitations overlay closely, we will observe strong coupling under realistic ambient conditions. 
If we describe however a single molecule undergoing the chemical reaction within a rather short time-frame, cavity and matter will undergo only few oscillations such that all resonant features will be washed-out by the short observation-time during a single reaction. This effect is further discussed in reference \cite{schaefer2022}. Consider furthermore that the vibrational modes will change during the reaction, a resonance is therefore only well defined for (meta)stable configurations. In contrast, if the vast majority of the molecules remain in its equilibrium configuration and the strong coupling exerts an effective force on the single molecule undergoing the reaction, then the resonant condition to modify chemical reactivity should be largely determined by the original resonance condition of the collective coupling. We would therefore intuitively expect single-molecular simulations to exhibit a much less sensitive resonant condition.

Apart from these general collectivity aspects for cavity mediated reactions,  the role of quantum (!) collectivity is another disputed theoretical question, i.e. to what extend a coherent multi-molecular polaritonic quantum state is formed and what are its implications on polaritonic reaction rates? In the subsequent argument, we will address the fundamental aspects of quantum collectivity and resonances for ground-state polaritonic reactions.  

\subsubsection{The role of collectivity in vibrational strong coupling and its local impact on the molecular potential energy surfaces}

A common opinion in the field of polaritonic chemistry is that there are two main contributors  to the observed changes in chemical reactions: Collectivity and quantumness, i.e., the emergence of coherent quantum states involving a large amount of molecules. This assumption is typically reflected by the choice of the model for the light-matter coupling, which is commonly a variant of the Dicke or Tavis-Cummings model. 
They are designed to provide the hybridization between light and the collective matter excitation. These models implicitly assume that there is a quantum coherence among a very large amount of molecules, which persists even at standard ambient conditions prevalent for typical chemical reactions. Widely used values for the number of coherently coupled molecules $N_{mol}$ vary between $10^{6}$ and $10^{11}$, suggesting quantum coherence over a mesoscopic length scale for a large number of molecules.~\cite{galego2015cavity,martinez2018can} Leaving aside the issue with creating coherent quantum states at a sizeable temperature and in solvation,~\cite{frowis2018macroscopic} we can scrutinize this basic assumption of quantum-coherence of a large amount of molecules by the parent Pauli-Fierz theory.

It is standard to derive the Dicke-type models starting from Eq.~\eqref{eq:pf_dip_h}. We therefore take the above Hamiltonian to describe the full ensemble of $N_{mol}$ molecules and make the usual assumption that we can describe the cavity by one effective mode. If we then make the further assumption that the individual molecules are far apart (dilute gas limit with non-overlapping electronic structure), and the coupling of the photon mode with frequency $\omega$ is weak for each individual molecule, we can find~\cite{ebbesen2016hybrid, schafer2018ab} that the Rabi splitting of the lowest polaritonic states is (in atomic units)
\begin{align}\label{eq:RabiSplit}
    \Omega \approx \sqrt{N_{mol}}\sqrt{\frac{8 \pi \omega}{L A}}|\braket{e}{\hat{\boldsymbol{X}}\cdot \boldsymbol{\epsilon}| g}, 
\end{align}
under resonant condition $\omega=\Delta\varepsilon_{ge}$, for the energy difference between the ground $g$ and excited $e$ state of a single of these (all identical) molecules.
Here we have used that the coupling vector $\boldsymbol{\lambda}$ between light and matter is determined by the polarization of that mode $\boldsymbol{\epsilon}$ and the coupling strength $|\boldsymbol{\lambda}| = \sqrt{4 \pi/(L A)}$, where $L$ is the length of the cavity and $A$ the surface corresponding to the mode volume~\cite{ebbesen2016hybrid, schafer2018ab}. Furthermore, the dipole operator of a single (all identical) molecule is denoted as $\hat{\boldsymbol{X}}$. Let us next take parameters from the experiment.\cite{thomas2016ground}  The resonantly coupled mode is $\nu = 860 cm^{-1}$, which with a simple model of the planar cavity of length $L=5.813 \mu m$
\begin{align}
    \nu = \frac{m}{2 n L},
\end{align}
leads to a refractive index $n=2$ (for the filled cavity) with mode number $m=2$. We note that the empty cavity has a smaller index of refraction $n\approx 1.4$ and thus the subsequently coupled mode has a higher wave number of about 1200 $cm^{-1}$.\cite{thomas2016ground} Therefore, in our investigated setting, the different refractive indices shift the cavity modes towards smaller wave numbers for the filled cavity. For the interpretation of the experimental data, one usually considers the values of the filled cavity. Using further the observed Rabi split $\Delta\nu = 98 cm^{-1}$  
at the above resonance frequency in combination with a conservative (i.e. large) estimate for the vibrational transition dipole element $|\braket{e}{\hat{\boldsymbol{X}}\cdot \boldsymbol{\epsilon}| g}| \approx 1$ [a.u.] at 860 $cm^{-1}$, derived from first-principle simulations,\cite{schafer2021shining} we find, with the standard choice $A=L^2$,  that $N_{mol} \approx 10^9$ is in accordance with the literature.

The question that now arises is whether all the assumptions made so far are justified or not, i.e. if we effectively have $N_{mol}$ quantum-coherently coupled molecules in the experiment. From the parent Pauli-Fierz Hamiltonian we can deduce further consequences of this widely taken assumption. The simplest one is found if we consider the unitarily equivalent velocity form of the Pauli-Fierz Hamiltonian of Eq.~\eqref{eq:pf_dip_h}. It is straightforward to calculate the depolarization (diamagnetic) shift of the empty cavity due to having $N_{mol}$ molecules quantum-coherently coupled inside the cavity, which gives in atomic units,~\cite{PhysRevLett.105.196402,rokaj2020free}
\begin{align}
    \omega_{d}^2 = \frac{4 \pi N_{mol}}{A L} \left(n_j + \sum_{i=1}^{N_j} \frac{Z_i^2}{M_i}\right),
\end{align}
where $n_j$ is the number of electrons of the individual molecule and $N_j$ the number of nuclei of the same molecule. Notice that collective light-matter coupling modifies the diamagnetic shift as well as the Rabi splitting, and both effects can be measured experimentally e.g., Ref.~\citenum{PhysRevLett.105.196402} for the depolarization shift.  If we just count the number of charges per molecule, we obtain roughly $(n_j + \sum_{i=1}^{N_j} Z_i^2/M_i) \approx 100$, which leads to a relative depolarization shift of,
\begin{align}
    \frac{\omega_d^2}{\omega^2} = \frac{4 \pi N_{mol} L^2}{L A \pi^2 c^2} 100 > 100,
\end{align}
independently of the chosen cavity surface $A$ (if we substitute Eq.~\eqref{eq:RabiSplit} to express $N_{mol}$). This would mean that the cavity frequency is blue-shifted to many multiples of the original frequency $\tilde{\omega}^2 = \omega^2 + \omega_d^2$, for collective coupling of $N_{mol}$ molecules. This would imply that a pure quantum effect  dominates over the classical shift towards smaller frequencies, due to the increased refractive index of the filled cavity. This shifting towards higher frequencies by multiples of the fundamental one is clearly not observed in experiment. From this result we can conclude that taking the assumption of a mesoscopic amount of  quantum-coherently coupled molecules  leads to fundamental inconsistencies, which are in clear disagreement with experimental observations. Notice that our consistency check does neither rule out quantum effects nor collective effects, but it objects to an overly simplistic combination of both. 

Let us therefore see next, whether \textit{ab initio} theory could shed some light on the issue of collective and quantum effect in ensembles of molecules.
Indeed, accurate coupled cluster calculations for the Hamiltonian of Eq.~\eqref{eq:pf_dip_h} recently showed that sizable collective effects can already emerge in the ground-state of molecular ensembles.\cite{HauglandJChemPhys2021} The analysis
performed on a cluster of water molecules demonstrated that QED induces non-additive contributions to the energy of the complex. In more detail, electron-photon correlation generates an energy stabilization that increase with the square of the number of molecules involved. Note that if the number of coherently-coupled molecules would increase to the mesoscopic scale, as anticipated by the Dicke-model results, we would observe a strong depolarization shift of the cavity frequency also in this situation, which is not the case in present experiments.

Moreover, linear response QEDFT reveals that for electronic strong coupling  local modifications of the electronic structure emerge in the vicinity of impurities, embedded within a collectively coupled  environment.\cite{sidler2020polaritonic,schaefer2022} This effect is also anticipated for vibrational strong coupling, which can  be described  by \textit{ab initio} linear response theory in a similar fashion.\cite{bonini2021ab} However, for the moment the computational verification remains an open research question. %The collective surrounding is explicitly represented by $N-1$ identically aligned nitrogen dimers with fixed (!) nuclear position.\cite{sidler2020polaritonic} This setting represents the closest possible \textit{ab initio} fit to a Dicke model, i.e. a coherent collective environment is initially imposed, and nicely reproduces all the spectral features of this model. \
The investigated environment is represented by $N-1$ identically aligned nitrogen dimers with fixed nuclear positions and $1.32$ nm separation. It describes a possible \textit{ab initio} realization of the Dicke model and accurately recovers collective bright and dark excitations.\cite{sidler2020polaritonic}
Within this setting, however, single-molecule strong coupling can emerge at the impurity due to the collective strong coupling of the environment to the cavity. In more detail, the \textit{ab initio} realization does not restrict the form of the dipolar excitation for each molecule, i.e. they are not enforced to be identical. Therefore, ultimately it is found that the environment of $N-1$ molecules can amplify the local oscillator strength, resulting  in an effectively amplified light-matter excitation of the impurity. Importantly, the local changes at the impurity are induced due to a strongly frequency-dependent polarization of the collective dipoles, which does not necessitate photonic quantum effects (i.e. it is a semi-classical effect). On the one hand, such single-molecule strong coupling embedded in an otherwise mostly classical ensemble could circumvent the above inconsistency arising for mesoscopic quantum states. On the other hand, it could also point towards the fact that we need to go beyond dipolar-coupling for the building of polaritonic models. Furthermore, it suggests that polaritonic modifications of the free-energy landscape are indeed expected to occur in experimental setups, while theoretical studies that suggest the opposite,\cite{li2020origin} may feature too restrictive theoretical assumptions  to be generally applicable (e.g. non-interacting molecules and bosonic Hartree-product Ansatz for the fermionic electronic structure).

Overall, the observation of this complex interplay suggests a paradigmatic shift in the understanding of polaritonic chemistry, which (partially) re-introduces the principle of locality for polaritons, a principle prevalent to describe chemical reactions (i.e. charge transfer). % of excited states (e.g. for photo-chemical reactions).
However, our latest \textit{ab initio} simulations cannot yet rationalize conditions under which a coherent collective environment can emerge and to what extend quantum or classical polarization effect play a role at ambient conditions. Moreover, we cannot yet disentangle the relevance of locally induced modifications vs. density of states (DOS) effects that emerge from populating dark states.\cite{xiang2019state} These and particularly the role of dark states\cite{du2021can} are important theoretical research questions, which should be addressed in future work using rigorous \textit{ab initio} methods.  Our hitherto existing simulations only reveal that collective effects induce local modifications, which can affect the free-energy landscape of a polaritonic ensemble and thus can be utilized to steer chemical reactivity.

 %We will comment later on in Sec.~... to go beyond. 

\subsubsection{Semi-classical non-equilibrium contributions to cavity mediated reaction rates under vibrational strong coupling}

After having considered modifications of the single-molecule strong coupling potential energy surfaces, we next focus on cavity induced dynamic effects, which we consider as the second key ingredient to rationalize cavity-mediated reaction rates. 

Before we start, we want to highlight that for the subsequent considerations, we assume that the entire polaritonic system is in thermal equilibrium at ambient conditions, which can \textit{a-priori} be described by the canonical density operator $\hat{\rho}(\hat{\bold{p}},\hat{\bold{P}},\hat{\bold{p}}_\alpha,\hat{\bold{r}},\hat{\bold{R}},\hat{\bold{q}}_\alpha)=\exp(-\hat{H}/k_B T)/\mathcal{Z}$, with $\mathcal{Z}=\text{tr}(\exp(-\hat{H}/k_B T))$ the corresponding partition function. We therefore rely on a temperature reservoir (instead of laser driving) as an external agent that populates (vibrationally) excited states. Hence, we deal with a thermalised distribution of excited states for the entire polaritonic system, in contrast to the non-thermal distribution induced by external laser. Having excited states populated is a necessary condition for the emergence of resonance effects. Notice that cavity modified reaction rates can be measured by means of mass spectrometry, i.e. in absence of any IR illumination,\cite{thomas2016ground} which leaves only thermal fluctuations as a source for vibrational excitations.  
In practice, the full quantum statistical treatment of realistic polaritonic system is computationally intractable, which we will try to circumvent by our subsequently developed argument based on established methods in molecular dynamics.
In essence, we will use that a reduced (!) canonical density operator, i.e. when tracing out some degrees of freedom, is not canonical anymore, unless the traced out degrees do not interact with the rest of the system (as commonly assumed for the main reaction coordinate in transition state theory). This can have interesting consequences for the dynamics of the reduced system (e.g. nuclear degrees of freedom in our polaritonic setting).
As stated before, dynamical effects, i.e. redistribution of kinetic energy, are considered the main driver of the experimentally observed resonance property for polaritonic reaction rates,\cite{thomas2016ground,schafer2021shining} which can already emerge in an entirely classical setup.\cite{wang2021cavitymodified}  

In the next step, we attempt to rationalize these recent theoretical findings further and develop an  alternative theoretical perspective to the prevalent quantum-collective point of view in polaritonic chemistry. For this purpose, we investigate vibrational strong coupling from the theoretical perspective of \textit{ab initio} MD,\cite{marx2009ab} which has been a reliable tool for decades to describe equilibrium nuclear dynamics in complex chemical setups. Therefore, we subsequently assume a ground-state chemical reaction (in accordance to the interpretation of the experiment\cite{thomas2016ground}) and stay on the lowest cavity Born-Oppenheimer (CBO)~\cite{flick2017atoms} surface. The CBO has repeatedly demonstrated to yield excellent results for vibrational strong coupling of isolated systems under NVE conditions (i.e. for constant particle number $N$, volume $V$ and energy $E$).\cite{flick2017cavity} In particular, the CBO ansatz assumes that the system under study can be partitioned into fast (electrons) and slow (nuclei, displacement fields) degrees of freedom. The fast degrees of freedom are treated quantum mechanically, which  depend parametrically on the slow degrees of freedom. Notice, that we will not restrict the ensemble size in our argument, which therefore allows for classical (not quantum) collective effects to emerge. 
The CBO Hamiltonian of the slow degrees of freedom, can be written in its simplest form (single mode, neglecting non-adiabatic couplings) as
\begin{eqnarray}
\hat{H}_{CBO}&:=&\sum_i^N \frac{\hat{\bold{P}}_i^2}{2M_i}+\frac{\hat{p}_\alpha^2}{2}+V_{CBO}(\bold{R},q_\alpha)\\
V_{CBO,gs}(\bold{R},q_\alpha)&:=&\sum_{i<j}^N\frac{e^2 Z_i Z_j}{|\bold{R}_i-\bold{R}_j|}
 +\frac{\omega_\alpha^2}{2}\Big(q_\alpha-\frac{\boldsymbol{\lambda}_\alpha}{\omega_\alpha}\cdot \bold{X}_R\Big)^2+\epsilon_{gs}\\%(\bold{R},q_\alpha)\\
 \epsilon_{gs}(\bold{R},q_\alpha)&:=&\bra{\psi_0(\bold{R},q_\alpha) }\hat{H}_e(\bold{R},q_\alpha)\ket{\psi_0(\bold{R},q_\alpha)}\label{eq:electronic_en}\\
 \hat{H}_e&:=&\sum_i^n \frac{\hat{\bold{p}}_i^2}{2m}+\sum_{i<j}^n\frac{e^2}{|\hat{\bold{r}}_i-\hat{\bold{r}}_j|}-\sum_{i,j}^{n,N}\frac{e^2 Z_j}{|\hat{\bold{r}}_i-\bold{R}_j|}\\
&&
 +\frac{1}{2}(\boldsymbol{\lambda}_\alpha\cdot \hat{\bold{X}}_r)^2+\boldsymbol{\lambda}_\alpha^2\hat{\bold{X}}_r\bold{X}_R-\omega_\alpha\boldsymbol{\lambda}_\alpha \cdot \hat{\bold{X}}_r q_\alpha\nonumber\\
 \bold{X}_R&:=&\sum_i^N Z_i e \bold{R}_i,\ \ \hat{\bold{X}}_r:=-\sum_i^n e \hat{\bold{r}}_i,
\end{eqnarray}
where the $\epsilon_{gs}$ denotes the minimized electronic ground state contribution to the potential energy surface (PES) and $\hat{H}_e $ indicates the parametrized electronic Hamiltonian operator with corresponding groundstate electronic eigenfunction $\psi_0$.

A next common assumption in MD,\cite{hutter2012car} in agreement with the usual transition-state theory in chemistry (e.g. Eyring\cite{eyring1935activated} or Marcus\cite{marcus1956theory,marcus1964chemical} theory), is to treat the ''slow'' degrees of freedom classically (typically nuclei). The resulting classical equation of motions give rise to conservative Hamiltonian dynamics under NVE conditions
\begin{eqnarray}
M_i\ddot{\bold{R}}_i&=&-\vec{\nabla}_i V_{CBO,gs}(\bold{R},q_\alpha)\label{eq:mnve}\\
\ddot{ q}_\alpha&=&-\frac{d}{dq_\alpha} V_{CBO,gs}(\bold{R},q_\alpha).\label{eq:ptnve}
\end{eqnarray}
To take into account the thermal bath that is present in room-temperature cavity experiments, we couple our system to a stochastic bath by means of Langevin equations of motion. This is a common standard in MD simulations\cite{brunger1984stochastic,schlick2010molecular,hutter2012car} that empirically accounts for environmental effects or additional degrees of freedom on the explicitly treated system, which typically gives rise to canonical equilibrium dynamics. Notice that the choice of the Langevin equations of motions is \textit{a priori} not motivated by cavity losses in our approach. 
%Because the experimentally\cite{?} observed resonance phenomena occur in thermal equilibrium i.e. a large amount of degrees of freedom is involved, it is common practice to model these environmental effects on the reaction coordinate stochastically by means of Langevin equations of motions.\cite{?} 
In more detail, we couple our dynamical system in Eqs.~\eqref{eq:mnve} and \eqref{eq:ptnve} to a stochastic bath 
%in thermal equilibrium (e.g. solvent environment or other not explicitly treated degrees of freedom). The stochastic bath shall
that exerts random forces and drag on the classical degrees arising from random collisions, %A standard way to account for the presence of these additional degrees of freedom, is by means of stochastic Langevin equations of motion, which transforms our Eqs. (), (), to,
\begin{eqnarray}
M_i\ddot{\bold{R}}_i&=&-\vec{\nabla}_i V_{CBO,gs}(\bold{R},q_\alpha)-\gamma M_i \dot{\bold{R}}\label{eq:langevin_m}\\
&&+\sqrt{2 M_i \gamma k_B T} \bold{S}\nonumber\\
\ddot{ q}_\alpha&=&-\frac{d}{dq_\alpha} V_{CBO,gs}(\bold{R},q_\alpha) -\gamma' \dot{q}_{\alpha}\label{eq:langevin_pt} \\
&&+\sqrt{2 \gamma' k_B T} S'\nonumber\\
\langle S(t)\rangle&=&0=\langle S'(t)\rangle \label{eq:stochmean}\\
\langle S(t)S(t^\prime)\rangle&=&\delta(t-t^\prime) = \langle S'(t)S'(t^\prime)\rangle\label{eq:stochdelta}
\end{eqnarray}
In the Langevin equation of motion a damping constant $\gamma$ was introduced which defines the velocity $\dot{\bold{R}}$ dependent friction term and $\bold{S}(t)$ corresponds to $d N$ independent stationary Gaussian processes with zero mean assuming a $d$-dimensional Euclidean space and accordingly for the displacement coordinate. Notice that Eqs. (\ref{eq:stochmean}) and (\ref{eq:stochdelta}) apply component-wise. 
The beauty of Langevin equation of motion is that for conservative forces it gives rise to the unique invariant Boltzmann distribution. For example, in absence of light-matter coupling ($\lambda_\alpha =0$) and for finite damping,  it is well-established that Eq.~(\ref{eq:langevin_m}) ensures the probability distribution $\rho_T(\bold{R},\bold{P})\propto \exp(-H_{CBO}(\lambda_\alpha=0)/k_B T)$,\cite{sachs2017langevin} due to the strictly conservative evolution on the PES given in Eq (\ref{eq:mnve}). Consequently, the matter system exposed to the stochastic bath obeys canonical equilibrium conditions at constant temperature $T$, which is independent of the chosen damping constant $\gamma$. In other words, provided that the dynamic evolution has explored the relevant phase space sufficiently, one can now infer  equilibrium properties for ergodic systems in the thermodynamic limit (e.g. transition rates). 

Now, the question arises how to deal with the additional displacement degree of freedom. From a mathematical perspective, adding the same Langevin bath of identical temperature with some damping $\gamma^\prime\neq0$ would automatically ensure a classical canonical equilibrium distribution for the coupled ($\lambda_\alpha>0$) degrees of freedom $\bold{R},q_\alpha$. Therefore, we would have recovered canonical equilibrium for our reduced polaritonic system (in nuclear and displacement field coordinates) as commonly assumed\cite{luk2017multiscale} and achieved\cite{li2020cavity} in the literature. Consequently, one expects~\cite{lelievre_stoltz_2016,sachs2017langevin} that the dependence on the internal parameters, such as the cavity frequency, is rather smooth, due to the fact that the whole system is conservative. Hence, for most observables no clear resonance with respect to changing the cavity frequency is expected. This feature has been demonstrated by various authors in the setting of transition-state theory applied to the polaritonic setting.~\cite{campos2020polaritonic,li2020origin} Their conclusion was that due to the theoretical absence of a resonance condition, either the semi-classical description was erroneous or that the experimentally observed effect is due to a different aspect (see list in Sec.~\ref{sec:list}). However, as initially stated, there is no guarantee that our reduced degrees of freedom obey canonical equilibrium dynamics.

Indeed, an immediate theoretical problem arises in the above semi-classical Langevin description under canonical equilibrium, which we are going to scrutinize below and provide an alternative framework for a thermalised polaritonic system: 
\begin{itemize}
    \item Canonical equilibrium implies that each classical degree of freedom possesses a kinetic energy of $k_B T/2$, i.e. $\langle\dot{q}_\alpha^2/2\rangle_T=k_B T/2$ due to the equipartition theorem. This ensures that the average velocity of each degree of freedom solely depends on its mass and temperature, but not on the potential energy. This has important consequences for the classical representation of the displacement degree of freedom, which now evolves at least three orders of magnitude faster than a Hydrogen atom. That is, the displacement field cannot be considered a slow degree of freedom anymore, which is needed to justify a classical thermal description! This contradiction also emerges for the simple quantum harmonic oscillator, i.e., the uncoupled photon degree of freedom. In that case we can solve the classical and the quantum thermal ensembles analytically. Both ensembles approximately agree for $k_B T \gg \omega_{\alpha}$, but not in our case, where $k_B T \lesssim \omega_{\alpha}$.
\end{itemize}

%\newpage
To resolve this theoretical equipartition issue for a coupled electron, ion and photon system, one could either treat the displacement field and the nuclei quantum statistically, which is, however, very challenging to do in practice, or we modify the Langevin equations to account for the quantum nature of the cavity photon fluctuations. In this regard, various possibilities arise such as adapting the distribution of the stochastic noise or introducing a different (effective) temperature for the displacement coordinate.~\cite{grosberg2015nonequilibrium, wang2020three} However, it is important to note that the physical photon field (and its fluctuations) is not determined by $\hat{q}_{\alpha}$ alone, but it is given by 
\begin{eqnarray}
\hat{E}_{\perp} &=& \boldsymbol{\lambda}_{\alpha} \omega_{\alpha} \hat{q}_{\alpha} -  \boldsymbol{\lambda}_{\alpha}  (\boldsymbol{\lambda}_{\alpha}\cdot\hat{\bold{X}}_{R}) + \boldsymbol{\lambda}_{\alpha}  (\boldsymbol{\lambda}_{\alpha}\cdot\hat{\bold{X}}_r).\label{eq:e_trans}
\end{eqnarray} 
This connection is of paramount importance to guarantee the physical condition that the transverse electric field $\hat{E}_{\perp}$ is zero for the entire polaritonic ensemble, i.e. $
\langle \hat{E}_\perp\rangle_T = \sum_k\bra*{k}\hat{E}_\perp\exp(-\hat{H}/k_B T)/\mathcal{Z} \ket*{k}=0$.\footnote{The simplest way to enforce the zero transverse electric field condition in our classical setting would be  an adiabatic assumption of the form $q_\alpha=\frac{\boldsymbol{\lambda}_\alpha \bold{X}}{\omega_\alpha}$.~\cite{flick2018cavity,schafer2018ab,schafer2021shining,schaefer2020relevance} However, in practice this might be a too severe restriction to recover the details of a cavity-modified chemical reaction.} Therefore, in our representation the photonic fluctuations are not only determined by the displacement field, but also by the fluctuations of the nuclear and even the electronic degrees of freedom.  

In contrast to the displacement-field fluctuations,  the classical equations of motions of the displacement fields themselves agree very well with the expectation value for the quantum equations of motions $\langle \hat{q}_{\alpha}(t) \rangle$, $\langle \hat{p}_{\alpha}(t) \rangle$, as long as we do not reach the (single-molecule) ultra-strong coupling regime.\cite{flick2019light,schafer2021making} The reason for this is found in the uncoupled photonic degrees of freedom, where by construction, irrespective of the initial state of the system,~\cite{spohn2004dynamics} the classical equation of motions reproduce exactly the expectation value of the quantum equations of motions. Hence, a classical description of our displacement field still seems appropriate, whereas a proper description of the displacement-field fluctuations would require considerable adaptation of the Langevin approach. For example, in an open quantum systems setting,  one could try to derive a Caldeira-Leggett\cite{caldeira1983path}-type of approximation, starting from the quantum master equations,\cite{breuer2002theory} which should explicitly account for the strong coupling conditions within the cavity. Consequently, quantum induced time-correlation effects would be expected in a more refined stochastic description. % which are not yet captured in our simple Langevin approach. 
In our Langevin setting, a computationally simple approximation arises from Eq. (\ref{eq:e_trans}) by assuming that, under vibrational strong coupling, the (fast) fluctuations of the displacement field are cancelled by the (fast) fluctuations of the electrons, while the fluctuations of the physical electromagnetic field $\hat{E}_{\perp}$ are dominated by the (slow) thermal fluctuations of the nuclei.
This assumed cancellation effect of fast fluctuations can simply be achieved by setting $\gamma^\prime=0$ in our Langevin setup in Eq. (\ref{eq:langevin_pt}), which automatically implies that the fluctuations of the physical field $\hat{E}_{\perp}$ are entirely driven by nuclear dipole fluctuations. In this case multiple stationary solutions for the probability-density function might arise and the zero transverse field condition might become important to single out the physical one.

\begin{figure}
\includegraphics[width=0.5\textwidth]{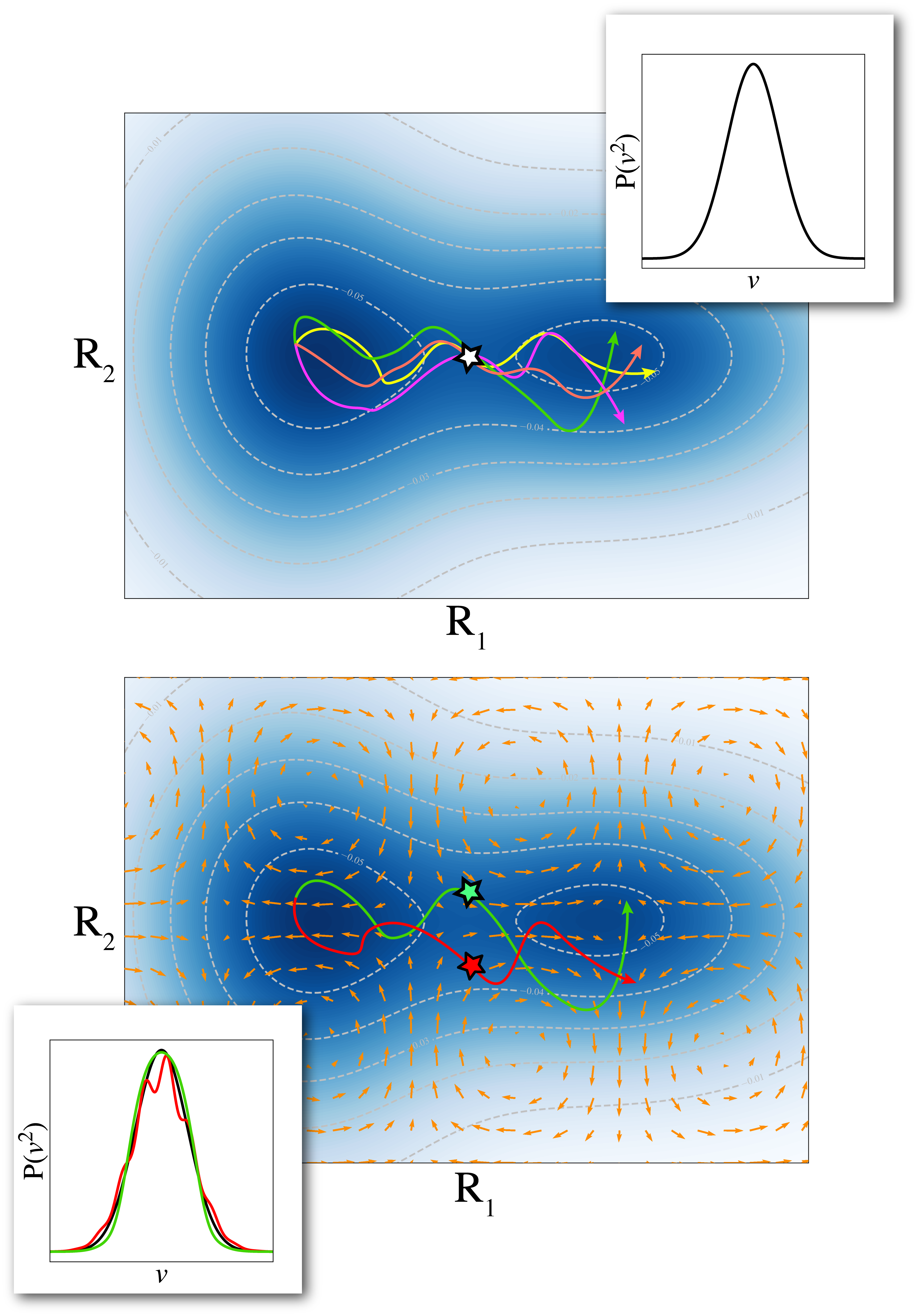}% Here is how to import EPS art
\caption{\label{fig:pes} Illustrative sketch  of different trajectories evolving on a double-well potential energy surface (blue sinks with grey isosurfaces) in canonical equilibrium (top) vs. stationary non-equilibrium dynamics (bottom). Units are chosen arbitrarily. Notice that the probability density $P(v^2)$ of each velocity degree of freedom is normally distributed in canonical equilibrium, where the temperature is related to its variance. In contrast, the emergence of (time-dependent) non-conservative forces (orange arrows) modifies the physical properties fundamentally, when coupled to a thermal bath. In that case, the stationary probability densities can deviate considerably from the Boltzmann solution. Moreover, one cannot necessarily identify relevant transition states (green and red star) from saddle points of the potential energy surface (white star). All of which effects could be relevant for the theoretical description of polaritonic reaction rates under vibrational strong coupling.}  
\end{figure}

Indeed, restricting our classical stochastic scattering events to the nuclei has astonishing consequences, since it introduces a time-dependent force component, acting as a constraint on the stochastic treatment of the nuclear degrees of freedom.  In more detail, the (now) deterministic photon degree of freedom connects $\bold{R}$ and $q_\alpha$ in a non-trivial way (see Eq.~(\ref{eq:langevin_pt}) for $\gamma^\prime=0$), which  violates the conservative-force assumption of the nuclear Langevin Eq.~(\ref{eq:langevin_m}). The non-conservative force entering the stochastic equations of motion will give rise to non-equilibrium nuclear dynamics for the nuclei,\cite{sachs2017langevin} exactly as we intended and visualize in Fig. \ref{fig:pes}.
Certainly, the emergence of non-conservative forces is somehow expected on physical grounds for a reduced polaritonic system, due to the coupling to the \textit{transversal} photonic fields. In that sense, our MD inspired approach ensures that effectively transversal force components are considered in the stochastic treatment of the classical nuclei dynamics. However, at the same time, our model preserves standard canonical equilibrium dynamics in the limiting case of zero coupling strength ($\lambda\rightarrow 0$), i.e. in absence of light-matter interaction, as one would expect. %However,  this important aspect vanishes as soon as the nuclear motion is treated classically by means of PES coupled to . %typically hidden in the quantum degrees of freedom if the system is partitioned into classical and quantum parts, i.e. when introducing potential energy surfaces, and its impact on the slow (classical) degrees of freedom is neglected. except in 
Consequently, our model provides a simple classical alternative to the full quantum-statistical treatment of the entire polaritonic system, which is practically unfeasible for realistic systems. In addition, our approach further rationalises the \textit{ab initio} QEDFT simulations in Ref.~\citenum{schafer2021shining} which observe a clear resonant condition in agreement with experiment and infer non-equilibrium nuclear dynamics under NVE conditions by explicitly considering  multiple nuclear degrees of freedoms. Note that in accordance with the QEDFT simulations, our semi-classical reasoning is restricted to a certain set of fundamental observables. In our case those are the nuclear coordinates $\bold{R}$, whereas predictions for fluctuations and other observables are less reliable. We further note that this makes the proposed classical probability distribution an auxiliary quantity analogous to the Kohn-Sham wave function in density-functional theories.~\cite{Evans_2016}

%In contrast to a full quantum statistical treatment of , our proposed Langevin equations of motion provide a simple classical alternative that explicitly accounts for cavity induced transversal forces under vibrational strong coupling. They are straightforward to implement in \textit{ab-initio} CBO molecular dynamics simulations. 
The proposed emergence of cavity-induced non-equilibrium nuclear dynamics under vibrational strong coupling could potentially explain why modified equilibrium rate theories were not able to reproduce the experimentally observed reaction rates based on reduced degrees of freedom (i.e., reaction coordinates). Indeed, the presence of non-conservative nuclear forces offers a tempting explanation to capture the observed resonance phenomena in an \textit{ab initio} MD setting, since one does not necessarily expect a smooth dependency on internal system parameters (e.g. cavity frequency) anymore. % in accordance to the thermal bath in the full quantum case. 
For example, it has been demonstrated that stochastic resonance phenomena can emerge in presence of non-conservative forces without (!) additional external driving, i.e. solely caused by stochastic noise.\cite{gang1993stochastic} It can be anticipated that the isolated polaritonic system indeed meets the necessary prerequisites if only the nuclei are weakly coupled to a Langevin bath. Therefore, looking at our initial reaction-rate mystery from the perspective of \textit{ab initio} MD, tuning the cavity on resonance could in fact mean that stochastic resonance conditions are met with respect to the thermal environment, which are then utilized to steer the (now) non-equilibrium nuclear reaction dynamics, whereas the entire polaritonic system remains in thermal equilibrium.

In any case, while this simple model  %towards a polaritonic reaction 
is inspired by \textit{ab initio} simulations that are in good agreement with experiment,\cite{schafer2021shining} we cannot further substantiate whether or not the approximations and assumptions involved  are sufficient to capture all the experimentally observed effects. For this, we particularly lack a detailed understanding of the thermal field fluctuations strongly coupled to matter, which are a crucial ingredient that eventually determines the exact non-canonical nature of the nuclear motion.   However, the proposed simple model already depicts that one can potentially realize the elusive resonance conditions for polaritonic reaction rates even in a simple semi-classical CBO perspective, unless tunneling becomes dominant (e.g. Ref. \citenum{galego2019cavity}). Moreover, our argument is in line with recently reported MD simulations that suggest a cavity enhanced relaxation rate (energy transfer) for selected, artificially heated molecules, which seems to affect mostly the tail of the energy-distribution, i.e. molecules that are likely to undergo a chemical reaction.\cite{li2021collective}
Overall, we think that the introduced model (based on \textit{ab initio} modelling) can serve as a computationally feasible starting point for further investigations on realistic chemical systems (e.g. involving explicit solvent molecules). It will help to unravel the origin and microscopic mechanism of photon-modified chemistry, which might pave the way towards the development of non-equilibrium reaction rate models that account for polaritonic resonance effects.

\section{Future perspective of ab initio polaritonic chemistry}
\label{sec:FutureQEDFT}

While the availability of \textit{ab initio} methodologies in polaritonic chemistry has already led to several surprising results and suggests a different (more local, semi-classical and non-equilibrium) perspective, there is still much to do to get a firm grasp of cavity-mediated chemistry. Among all of the mentioned aspects (see Sec.~\ref{sec:list}), the major open issues are the contributions of collective (possibly quantum) effects on chemical reactions at ambient conditions and the influence of the environment (openness of the cavity and solvation effects). While \textit{ab initio} simulations have already targeted several of these issues (collective effects~\cite{sidler2020polaritonic,schaefer2021shortcut,schaefer2022}, open cavities~\cite{jestadt2019light,schaefer2021shortcut},...), and the above proposed model approach can be straightforwardly extended to an ensemble of molecules and the inclusion of solvents, it is evident that further, more refined investigations are necessary.

In this context, QEDFT provides a highly versatile toolkit, since it can be used to simulate even the full minimal coupling problem of electrons, nuclei and photons, where the cavity is described on the same level of theory~\cite{jestadt2019light}.

The fundamental Hamiltonian of non-relativistic QED is the Pauli-Fierz Hamiltonian in Coulomb gauge given by~\cite{spohn2004dynamics, ruggenthaler2018quantum,jestadt2019light} 
%The recent theoretical breakthroughs confirm that the full \textit{ab initio} description of polaritonic chemistry based on the rigorous Pauli-Fierz Hamiltonian is crucial for the interpretation of experimental results. Moreover, they demonstrate that \textit{ab initio} simulation methods provide an invaluable tool to address open  questions in polaritonic chemistry. %However, up until now, the full potential of QEDFT has not been explored. 
%Motivated by the seminal results on cavity-mediated reaction dynamics, two fundamental research directions arise, which should be targeted by the future theoretical developments in polaritonic chemistry: 
%\subsection{Future developments in the \textit{ab-initio} description of polaritonic chemistry}
%QEDFT has been established as an invaluable tool to address open theoretical questions of polaritonic chemistry from first principles. 
%The rigorous theoretical foundations of QEDFT allows the systematic refinement of the underlying non-relativistic Pauli-Fierz Hamiltonian towards an even more general and accurate description than introduced in Eq. (\ref{eq:pf_dip_h}). This opens room for many novel discoveries. In the following we would like to shortly illustrate promising future development steps with potential applications. For this purpose the full minimal-coupling Hamiltonian in Coulomb gauge is introduced,
\begin{eqnarray}
\hat{H}(t)&=&\sum_{i=1}^n \frac{1}{2m}\bigg[\bold{\hat{\sigma}}_i\cdot \Big(-i \hbar \nabla_{\bold{r}_i}+\frac{e}{c} \hat{\bold{A}}_\perp (\bold{\hat{r}}_i,t)\Big)\bigg]^2\nonumber\\
&&+\sum_{i=1}^N\Bigg\{ \frac{1}{2M_i} \Big(-i \hbar \nabla_{\bold{\hat{R}}_i}-\frac{Z_i e}{c} \hat{\bold{A}}_\perp (\bold{\hat{R}}_i,t)\Big)^2\nonumber\\
&&-\frac{Z_i e \hbar}{2M_i c}\bold{\hat{S}}_i^{(s_i/2)}\cdot \Big(\nabla_{\bold{\hat{R}}_i}\times \hat{\bold{A}}_\perp (\bold{\hat{R}}_i,t)\Big)\Bigg\}\nonumber \\
&&+\sum_{i<j}^n\frac{e^2}{|\hat{\bold{r}}_i-\hat{\bold{r}}_j|}+\sum_{i<j}^N\frac{e^2 Z_i Z_j}{|\hat{\bold{R}}_i-\hat{\bold{R}}_j|}\nonumber\\
&&-\sum_{i,j}^{n,N}\frac{e^2 Z_j}{|\hat{\bold{r}}_i-\hat{\bold{R}}_j|}+\sum_{\bold{k},\lambda}\hbar \omega_k \hat{a}^\dagger_{\bold{k},\lambda} \hat{a}_{\bold{k},\lambda} 
\label{eq:pf_full_h}
\end{eqnarray}

Here, the transversal field operator $\hat{\bold{A}}_\perp (\bold{\hat{r}},t)$ is spatially dependent ($\bold{\hat{r}}$) and contains an explicit time-dependency $t$ that accounts for possible classical external driving. Electronic spin contributions are accounted for by the Pauli matrices $\hat{\bold{\sigma}}_i$, whereas nuclear spins are denoted by the vector of spin $s_i/2$ matrices $\bold{\hat{S}}_i^{(s_i/2)}$ with $s_i$ even/odd depending on the nuclear mass number.

The Pauli-Fierz Hamiltonian in full minimal coupling (see Eq.~\eqref{eq:pf_full_h}) is the low-energy limit of the relativistic QED Hamiltonian.~\cite{greiner2013field,ruggenthaler2014quantum} While it keeps the quantized photon field fully relativistic, it assumes that the charged particles have small kinetic energy such that the usual non-relativistic momentum operator is applicable. We note that in adding also the nuclei/ions as effective quantum particles, we go beyond the usual setting of QED which is defined for Dirac electrons only. While the Pauli-Fierz theory is mathematically similar to quantum mechanics and thus allows for uniquely defined wave functions,~\cite{spohn2004dynamics} the wave function of this quantum field theory is an numerically unfeasible object (besides the many particle degrees of freedom we have infinitely many photon degrees of freedom). Therefore one needs to use many-body methods that reformulate the Pauli-Fierz quantum-field theory in terms of reduced quantities.~\cite{ruggenthaler2014quantum, ruggenthaler2015ground,de2016unified,buchholz2019reduced,karlsson2021fast} The most developed of these approaches is QEDFT, where the wave function is replaced by the current density and the vector potential.~\cite{jestadt2019light} This substitution allows to recast the problem in terms of an auxiliary non-interacting system of electrons, nuclei and photons that generate the same densities and potentials. While in principle an exact reformulation of the full field theory, in practice the accuracy of a QEDFT simulation strongly depends on the approximations used for the effective fields and currents, which force the non-interacting system to reproduce the fully interacting one. One of the main advantages of QEDFT is that it seamlessly connects full minimal coupling to approximate version like the long-wavelength limit in the few mode approximation as given in Eq.~\eqref{eq:pf_dip_h}.~\cite{jestadt2019light} This provides the possibility of a systematic theoretical refinement of the \textit{ab initio} QED description of cavity-modified chemistry. 

With this highest level of theory we can (at least in principle) investigate all of the above listed aspects (see Sec.~\ref{sec:list}) in great detail, under various chemical setting, which allows to scrutinize the impact of common assumptions, such as the dipole approximation or to treat the electromagnetic field as an external perturbation only.~\cite{jestadt2019light} There are many situations, e.g. for the strongly debated super-radiant phase transition, where these type of aspects are assumed to be decisive.~\cite{PhysRevA.98.053819,PhysRevB.100.121109, rokaj2020free,andolina2020theory} How much they contribute to cavity-mediated chemical reactions has to be seen.

Besides investigating more realistic descriptions of polaritonic situations with QEDFT, there are further important theoretical topics that are actively explored. Among others this includes:
\begin{enumerate}
    \item \textbf{Polaritonic functionals:} 
    Similar to ordinary DFT, the success of QEDFT is determined by the availability of reliable and accurate approximate exchange-correlation functionals. So far available QEDFT functionals mostly base on perturbation theory for the light-matter interaction,~\cite{pellegrini2015optimized,flick2018ab,schaefer2020thesis,flick2021simple} whereas non-perturbative approaches depend on the use of polaritonic (higher-dimensional) constructions.~\cite{buchholz2019reduced, nielsen2018dressed, buchholz2020light}
    %as well as orbital dependent~\cite{pellegrini2015optimized,flick2018ab,schaefer2020thesis,flick2021simple} and only recently a first QEDFT functional based on the adiabatic-connection fluctuation-dissipation theorem has been presented~\cite{flick2021simple}. 
    %A first non-perturbative approach was based on polaritonic wave functions~\cite{buchholz2019reduced, nielsen2018dressed, buchholz2020light}. 
    However, recent developments~\cite{schafer2021making} suggest a new route based on effective photon-free Hamiltonians that have provided the first non-perturbative local-density like functional for QEDFT – allowing the self-consistent treatment of quantum light-matter interactions for sizeable systems.
    Nevertheless, it remains clear that considerable effort will be necessary in order to reach the same level of sophistication and versatility that has been established over decades for ground state DFT. To attain this goal, coupled-cluster theory based \textit{ab initio} QED methods\cite{haugland2020coupled,mordovina2020polaritonic} provide a valuable benchmark  for small systems, which is vital for the ongoing development of QEDFT.

    \item \textbf{Coulomb gauge in the long wavelength approximation:} In accordance to the above development of QEDFT functionals, often the \textit{ab initio} simulation in the Coulomb gauge with dipole approximation is preferable over the simulation in the unitarily equivalent length form. The fundamental advantage of the Coulomb gauge in long wavelength approximation (imposed on Eq. (\ref{eq:pf_full_h})) is twofold, compared with the standard length-gauge representation given in Eq.~(\ref{eq:pf_dip_h}). Since it is compatible with periodic boundary conditions on the matter system, and thus it's formulation is origin independent, ab initio simulations become feasible in a unified setting from the gaseous to solid (periodic) phase under strong light-matter interaction. This does not only allow the time-resolved study of strong-light matter interaction on critical phenomena, but it also provides a good starting point towards more realistic simulation setups accounting for explicit solvent molecules. Moreover, periodic boundary conditions are also a desirable feature for the future development of cavity \textit{ab initio}  MD methods yielding access to cavity-modified nuclear dynamics on long timescales under thermal equilibrium.
  
    \item \textbf{Classical external driving:} Combining time-dependent external driving with cavities opens a promising route towards the unprecedented control of molecular as well as material properties.\cite{hubener2021engineering} Within our QEDFT approach, classical external laser fields are straightforward to include, which can be employed to pump resonant photon modes as well as to modify matter properties. Here we would like to highlight one special feature of polaritonic systems. Collectively coupled polaritonic systems possess two different kind of excitations, i.e. \textit{bright} excitations, which respond to the external laser driving and \textit{dark} excitations that remain (virtually) unaffected, but can for example be populated thermally. This opens unique opportunities to utilize the complex interplay between thermal motion, external driving, resonance conditions and multiple modes to enter novel physical regimes and to spectroscopically probe polaritonic physics~\cite{lloyd20212021}
\end{enumerate}

\section{Conclusion}

In this work, we aimed to illustrate the benefit of \textit{ab initio} methods for the theoretical understanding of polaritonic chemistry. %, with their alternative focus. 
To our opinion, they offer a mostly unbiased approach to disentangle the vast complexity of polaritonic chemistry and to identify the most relevant underlying mechanisms. These aspects were exemplified with respect to the quantum collective paradigm and for the experimentally observed resonance conditions in polaritonic reactions under  vibrational strong coupling. 

Indeed, a fundamental theoretical contradiction was uncovered for  the quantum (!) collective coupling of a mesoscopic number of molecules, when considering the depolarization shift of the Pauli-Fierz Hamiltonian for the interpretation of experimental data, based on models from quantum optics. This suggests that the predominant theoretical interpretation of vibrational strong coupling, in terms of a collective quantum state on a mesoscopic scale, needs refinement at ambient conditions. 
Moreover, based on recently published QEDFT results,\cite{schafer2021shining,sidler2020polaritonic} a simple, but computationally efficient, Langevin perspective was introduced for the interpretation of the experimentally observed resonance phenomena in polaritonic reaction rates under vibrational strong coupling. The proposed semi-classical model has interesting features, since it allows for the emergence of cavity induced non-equilibrium nuclear dynamics, which then can give rise to (stochastic) resonance phenomena even in the absence of external periodic driving. 

To the authors' opinion, combining the knowledge from recent \textit{ab initio} simulations\cite{HauglandJChemPhys2021,sidler2020polaritonic,schafer2021shining} with the  aforementioned theoretical arguments, indeed  suggest a paradigmatic shift away from the prevailing, collective quantum interpretation on mesocopic scales,  towards a more local, non-equilibrium, (semi)-classically, driven mechanism for groundstate chemical reactions under (collective) vibrational strong coupling. 
Certainly,  careful validation against more rigorous \textit{ab initio} and quantum-statistical methods will be required to support our proposed model and to further substantiate our perspective on groundstate polaritonic reactions.

Aside from addressing the reaction rate mystery, future steps in the developments of ab initio methods with a focus on QEDFT were sketched, which involve novel polaritonic functionals to reach larger system sizes, periodic boundary conditions to access all states of matter from gaseous to solid, as well as the inclusion of external laser driving.
Overall, we believe that many future discoveries in polaritonic chemistry will emerge from these  developments, which eventually can be further rationalized into models, aiming for the intuitive understanding of polaritonic chemistry and polaritonic materials in general.

\begin{acknowledgments}
We thank G\"oran Johansson for critical comments and inspiring discussions.
This work was made possible through the support of the RouTe Project (13N14839), financed by the Federal Ministry of Education and Research (Bundesministerium für Bildung und Forschung (BMBF)) and supported by the European Research Council (ERC-2015-AdG694097), the Swedish Research Council (VR) through Grant No. 2016-06059, the Cluster of Excellence “CUI: Advanced Imaging of Matter” of the Deutsche Forschungsgemeinschaft (DFG), EXC 2056, project ID 390715994 and the Grupos Consolidados (IT1249-19). The Flatiron Institute is a division of the Simons Foundation.
\end{acknowledgments}

\section*{Data Availability Statement}
Data sharing is not applicable to this article as no new data were created or analyzed in this
study.

%\appendix

%\section{Appendixes}

\nocite{*}
\bibliography{perspective}% Produces the bibliography via BibTeX.

\end{document}